\documentclass{article} 

\usepackage[a4paper]{geometry}

\usepackage{amsmath}
\usepackage{authblk}
\usepackage{caption} 
\usepackage{subcaption} 
\usepackage{graphicx}

\usepackage[utf8]{inputenc}
\usepackage{multicol}
\usepackage{algpseudocode,algorithm,algorithmicx}
\usepackage{hyperref}
\begin{document}
\title{\textbf{A heuristic approach to the far-future state of a universe dominated by phantom energy} }

\author{Nikolaos Kalntis}

\affil{Ernest Orlando Lawrence Berkeley National Laboratory, University of California, Berkeley, CA 94720, USA, 
Email: \href{mailto:nkalntis@lbl.gov}{nkalntis@lbl.gov}}

\renewcommand\Affilfont{\itshape\small}
\date{}
\maketitle   

\begin{abstract}
This work is based on a cosmological scenario of a universe dominated by phantom energy with equation of state parameter $w<-1$ and the analysis of its asymptotic behaviour in the far-future. The author discusses whether a Big Rip singularity could be reached in the future. Working in the context of general relativity, it is argued that the Big Rip singularity could be avoided due to the gravitational Schwinger pair-production, even if no other particle-creating contribution takes place. In this model, the universe is described in its far-future by a state of a constant but large Hubble rate and energy density, as well as of a constant but low horizon entropy. Similar conditions existed at the beginning of the universe. Therefore, according to this analysis, not only the Big Rip singularity could be avoided in the far-future but also the universe could asymptotically be led to a new inflationary phase, after which more and more universes could be created.

\end{abstract}

\vspace{0.3cm} 

\begin{center}
\end{center}

\newpage

\section{Introduction}
\label{subsec:introduction}

The analysis of this paper starts with a brief review of the standard arguments that lead to the introduction of dark energy. The framework is a homogeneous, isotropic universe, which is spatially flat ($\kappa = 0$). Its spacetime can be described by the Friedmann-Lemaître-Robertson-Walker (FLRW) metric \footnote{In this paper, the author works in Planck units $\hbar$ = c = $k_B$ = 1.}

\begin{equation}
    ds^2 = - dt^2 + a(t)^2(dr^2 + r^2 d\Omega^2) \,,
\end{equation}

with a(t) the scale factor. Solving the Einstein equations $R_{\mu \nu} - \frac{1}{2} R g_{\mu \nu} = 8 \pi G T_{\mu \nu}$ for this type of metric leads to the well-known Friedmann equations

\begin{equation}
    H^2 = \Big(\frac{\Dot{a}}{a} \Big)^2 = \frac{8 \pi G}{3} \rho \label{fried1} \,,
\end{equation}

\begin{equation}
    \Dot{H} + H^2 = \frac{\ddot{a}}{a}  = - \frac{4 \pi G}{3}(\rho + 3 P) \label{fried2} \,.
\end{equation}

with $\rho$ the energy density and $P$ the pressure of the respective component. $P$ and $\rho$ follow an equation of state 

\begin{equation}
    P = w \rho \label{eos} \,,
\end{equation}

with $w$ the equation of state parameter. Combining \eqref{fried1} and \eqref{fried2} results in

\begin{equation}
    \dot{H} = - 4 \pi G (\rho + P) \label{Hdot} \,.
\end{equation}

Simultaneously, the relation for the energy-momentum conservation $\nabla_{\mu} T^{\mu \nu} = 0$ leads to the continuity equation

\begin{equation}
    \dot{\rho}  + 3 H (\rho + P) = 0 \label{emc}  \,. 
\end{equation}

The observation of the accelerating expansion of the universe has been one of the most remarkable discoveries in Astrophysics and Cosmology \cite{perlmutter1999}, \cite{riess1998}. This expansion cannot be explained by the current forms of matter and radiation known so far. Looking at the equation \eqref{fried2}, a form of energy with negative pressure and $w<-1/3$ should exist, so that $\ddot{a} > 0$. This substance, called dark energy, has the opposite effect (anti-gravitational) to that of gravity, accounts for almost 3/4 of the energy content of the universe and despite intense research since its discovery, it is yet unknown what this form of energy is \cite{planck_results_2018}.

One interesting case to consider is that of dark energy as a fluid with equation of state $w < -1$. Then the dark energy is called phantom energy. In this case, equations \eqref{Hdot} and \eqref{emc} lead to $\dot{H} > 0$ and $\dot{\rho} > 0$, i.e. the Hubble rate and the energy density increase with time. These models violate the so-called Null Energy Condition (NEC), which states that for any light-like vector $\eta^{\mu}$, with $g_{\mu \nu} \eta^{\mu} \eta^{\nu} = 0$, the following condition for the energy-momentum tensor should hold $T_{\mu\nu} \eta^{\mu} \eta^{\nu} \geq 0$. In the case of a spatially-flat FLRW universe this equivalently leads to the fact that $\rho + P \geq 0$, which is violated by a fluid with equation of state with $w<-1$. 

On the one hand, it is exactly the violation of NEC that has made a large part of the academic community not consider the models of dark energy with equation of state parameter $w<-1$ viable. This is mainly based on the grounds that the violation of NEC is proven to lead to instabilities (in the form of ``ghosts" and ``tachyons") for a large class of models \cite{rattazzi2006}. 

On the other hand, it has been proven that these instabilities can successfully be controlled in the context of effective field theories, despite the violation of NEC \cite{senatore2006}. In this work, it will be assumed that these instabilities are under control. Additionally, although there is the dominant belief in the scientific community that the dark energy ought to be the energy of the vacuum in the form of a cosmological constant $\Lambda$ or equivalently a substance with $w=-1$, it is still not clear from the data whether the dark energy fluid has $w$ greater, less or equal to $-1$ \cite{planck_results_2021}, \cite{dark_energy_survey}. Only future experiments have the potential to distinguish $w = -1$ from percent-level deviations. 

Therefore, the idea that dark energy could be phantom energy is still an open possibility that could be discovered in the future. It is definitely worth exploring the implications of this idea into more depth. Over the last years, there has been extensive analysis in this direction. The most interesting realisation is that if dark energy is indeed phantom energy then it could lead to the so-called “Big Rip" scenario, where eventually every part of the universe could be “ripped" apart in a finite amount of time due to the super-exponential expansion of the underlying space\footnote{In particular, the energy density and thus the Hubble rate reach infinite values in a finite amount of time.} \cite{big_rip}, \cite{big_rip_2}, \cite{big_rip_3}. However, infinities usually indicate an incompleteness of the respective theory in some specific limits. Therefore, a reasonable question to ask is whether the Big Rip singularity can be avoided in any way possible in the far-future of the universe.

\section{Far-future state of the universe}

\subsection{Gravitational Schwinger pair-production in the far-future of the universe}
\label{subsec:schwinger_pairs}

To answer the question posed in the previous section, one has to think of what could possibly act as a counterpart to the super-exponential expansion of the universe caused by the phantom energy \footnote{In this paper the author solely works in the context of general relativity, without assuming $f(R)$ theories of modified gravity (for a review on this topic, see \cite{fR_GR_review}, \cite{fR_review_2}). In this context, there have been various analysis of how the Big Rip or other type of singularities could be avoided in the far-future. Here, some of the analyses tackling of these topics are mentioned \cite{fR_big_rip}, \cite{fR_big_rip_2}, \cite{phantom_quantum_cosmology}, \cite{no_big_rip},  \cite{no_big_rip_2},  \cite{no_big_rip_3},  \cite{no_big_rip_4}, \cite{no_big_rip_5}, \cite{no_big_rip_6}, \cite{no_big_rip_7}, \cite{no_big_rip_8}, \cite{qg_1}, \cite{qg_2}, \cite{qg_3}, \cite{qg_4}.}. In this work, the attention is turned to a phenomenon called the Schwinger effect. This phenomenon, first derived by Julian Schwinger in 1951, is the production of a particle-antiparticle pair out of the vacuum in the presence of a strong electric field \cite{schwinger_effect}. The spectrum of the produced pairs is given by the formula \cite{schwinger_rev_1}, \cite{schwinger_rev_2}

\begin{equation}
    \frac{dn_{S}(p)}{d^3\Vec{p}} \approx \theta(p_z) \theta(qeEx^0 - p_z) \exp{\Big(-\frac{\pi(p_T^2+m^2)}{qeE}\Big)} \,, \label{ndensSchw} 
\end{equation}

with $p_T$ the transverse momentum, $x^0$ the time component of the spacetime 4-vector, $x^\mu = (x^0, \Vec{x})$, $m$ the mass, $qe$ the charge of each of the produced particles of the pair and $E$ the electric field.

It is easy to observe that one should create a very strong electric field, i.e. $qeE >> (p_T^2+m^2)$, in order for the exponential factor in equation \eqref{ndensSchw} not to suppress the pair production from the vacuum. Such large electric fields are difficult to be produced in the laboratories which is why the Schwinger effect has not yet been observed in Nature.  However, one may still try to think how this effect could potentially play a role on a cosmological scale. To make this connection, the horizon temperature $T_h$ for the case of de Sitter spacetime is defined \cite{hawking_radiation}, \cite{gibbons_hawking_1}, \cite{gibbons_hawking_2}

\begin{equation}
    T_h := \frac{\kappa_{h}}{2\pi} = \frac{H}{2\pi} \label{Hawking_temperature_dS}\,,
\end{equation}

where $\kappa_h$ is the acceleration of gravity at the surface of de Sitter horizon and is proportional to the Hubble rate H in the case of de Sitter spacetime. Connecting the relation \eqref{Hawking_temperature_dS} with the Unruh temperature of a uniformly accelerated observer \cite{unruh}, one can find the following relation between the gravitational acceleration $k_h$ of a particle of mass $m$, charge $qe$ and transverse momentum $p_{T}$ with the electric field $E$ \cite{Th_E}, \cite{castorina}\footnote{Given that one may equate the gravitational acceleration $k_h$ with the acceleration of the particle.}

\begin{equation}
    \kappa_h = \frac{qeE}{\sqrt{p_T^2+m^2}} \label{gravity_acceleration} \,.
\end{equation}

The combination of the equations \eqref{Hawking_temperature_dS} and \eqref{gravity_acceleration} results in

\begin{equation}
   qeE = 2\pi T_{h} \sqrt{p_T^2+m^2} = 2\pi H \sqrt{p_T^2+m^2} \label{qeE_H} \,.  
\end{equation}

One may calculate the contribution of the gravitational Schwinger effect to the energy density. According to the definition of the energy density, this is as following 

\begin{equation}
    \rho_{S} = \int d^3\Vec{p}~n_{S}(p) \sqrt{\Vec{p}^2 + m^2} \label{rho_S} \,,
\end{equation}

Then using the relations \eqref{ndensSchw} and \eqref{qeE_H}, \eqref{rho_S} becomes 

\begin{equation}
        \rho_S  = \int_{0}^{\infty} d(p_T^2)\int_{0}^{T_h \sqrt{p_T^2+m^2}x^0} dp_z \sqrt{p_T^2+p_z^2+m^2} \exp{\Big(- \frac{\sqrt{p_T^2+m^2}}{2T_{h}}\Big)} \label{rho_S_1} \,.
\end{equation}

Assuming light-produced particles ($m \approx 0$) \eqref{rho_S_1} becomes 

\begin{equation}
    \rho_S \approx C_S H^4 \label{rho_S_final} \,,
\end{equation}

with $C_S>0$ the respective coefficient. This is a radiation term, which is expected given the light-produced particles and whose energy density scales like $T^4$ according to the Stefan-Boltzmann Law. However, in the case of the gravitational production of particles, the coefficient does not have to be the same as in the Stefan-Boltzmann Law.\footnote{From now on, the gravitational Schwinger effect is simply referred to as Schwinger effect.} 

Considering the contributions of the phantom energy and the Schwinger effect, the Friedmann equation \eqref{fried1} becomes

\begin{equation}
    H^2 = \frac{8 \pi G}{3} (\rho_{DE} + \rho_S) =  \frac{8 \pi G}{3}(\rho_{DE} + C_S H^4 \label{H_2}) \,.
\end{equation}

Simultaneously, considering \eqref{rho_S_final} and that $\Dot{\rho}_{S}  = \displaystyle\frac{d\rho_S}{dt} \sim \displaystyle\frac{H^4}{H^{-1}} = H^5$, the continuity equation \eqref{emc} becomes

\begin{equation}
    \dot{\rho}_{DE} - 3 H |w_{DE} + 1| \rho_{DE} +  \Tilde{C}_S H^5 = 0 \label{continuity_eq_DE_source} \,,
\end{equation}

where it is assumed that $w_{DE} < -1$ is a constant. Also $\Tilde{C}_S$ is a constant function of $C_S$ and $w_{DE}$\footnote{For convenience, one may set $\Tilde{C}_S \equiv 3|w_{DE} + 1| C_S$.}. 

The analysis above has been done in the context of an almost de Sitter space, although $w_{DE}<-1$. This is a reasonable assumption because, even if $w_{DE}<-1$, still it should be close to $-1$, according to the observations, as discussed in paragraph \eqref{subsec:introduction}. Additionally, there is no concrete and universally accepted definition of the surface gravity for general curved spacetimes \cite{surface_gravity}, so $\eqref{Hawking_temperature_dS}$ could not be used for the general case of $w_{DE}<-1$. Also for our model we assume that there no catastrophic instabilities, which can be justified in the context of \cite{senatore2006}, as mentioned in the paragraph \eqref{subsec:introduction}. 

It is noted that the phantom fluid does not evolve independently from the Schwinger pair production, so the continuity equation \eqref{emc} does not hold independently for each component. One can see that the Schwinger effect contributes with a term proportional to $H^4$ and $H^5$ in \eqref{H_2} and \eqref{continuity_eq_DE_source} respectively. These terms become important to higher values of $H$, therefore later in the evolution of the universe if it is dominated by a phantom substance. Even though at later times the solutions of \eqref{H_2} and \eqref{continuity_eq_DE_source} start deviating from the de Sitter case, one expects gravitational particle production to occur whenever there is an event horizon (as with Hawking radiation in the case of back holes \cite{hawking_radiation}). Therefore, one would expect at least a radiation term proportional to $H^4$ and a source term proportional to $H^5$ in the Friedmann and continuity equations as in \eqref{H_2} and \eqref{continuity_eq_DE_source}, independent of whether the spacetime is described by a metric close to de Sitter or not. 

The goal now is to solve the equations \eqref{H_2} and \eqref{continuity_eq_DE_source} simultaneously. First of all, the equation \eqref{H_2} has two solutions in $H^2$

\begin{equation}
    H^2 = \frac{3}{16 \pi G C_S} \Big(1 \pm \sqrt{1-\frac{256 \pi^2 G^2 C_S}{9} \rho_{DE}} \Big) \label{H_2_2} \,.
\end{equation}

In \eqref{H_2_2} only the solution with the relative minus sign is kept, since this is the one that reduces to $H^2 \approx \displaystyle\frac{8 \pi G}{3} \rho_{DE}$ for small $H$, where the $H^4$ term coming from the Schwinger effect in \eqref{H_2} is negligible\footnote{The other solution  with the plus sign reduces to $H^{2} \approx \displaystyle\frac{3}{8 \pi G C_S} =$ constant in the same limit.}. Additionally, this solution has an upper value for $\rho_{DE}$, which is $\rho_{DE, max} = \displaystyle\frac{9}{256 \pi^2 G^2 C_S}$. This leads to a maximum value of $H^2$ which is $H^2_{max} = \displaystyle\frac{3}{16 \pi G C_S}$ \footnote{$H_{max}, \rho_{max} \rightarrow \infty $ as $C_S \rightarrow 0$, which is the Big Rip case if the Schwinger particle production would be negligible. Because this analysis is done only in the context of general relativity, without considering modifications of gravity in the UV regime, one should require that $\rho_{DE,max} << \rho_{Planck}$, and $H_{max} << M_{Planck}$, which equivalently means that $C_S >> 1$. }. Setting also $R_{DE} \equiv \rho_{DE}/\rho_{DE, max}$, the solution with the relative minus sign in equation \eqref{H_2_2} becomes 

\begin{equation}
    H^2 = H_{max}^2 (1-\sqrt{1-R_{DE}}) \label{H_2_final} \,.
\end{equation}

Using the equation \eqref{H_2_final}, the continuity equation \eqref{continuity_eq_DE_source} becomes

\begin{equation}
    \frac{dR_{DE}}{dt} = 6 |1+w_{DE}|H_{max}(1-\sqrt{1-R_{DE}})^{1/2}(R_{DE} + \sqrt{1-R_{DE}}-1) \label{R_diff} \,.
\end{equation}

Integrating the equation \eqref{R_diff} and using the fact that $R_{DE}(t_{max})=1$ results in

\begin{equation}
    \frac{\rho_{DE}}{\rho_{DE, max}} = 1 - \bigg[1- \frac{1}{(1+\frac{3}{2}|1+w_{DE}|H_{max}(t_{max}-t))^2} \bigg]^2  \label{R_DE_t} \,.
\end{equation}

Putting \eqref{R_DE_t} into \eqref{H_2_2} leads to

\begin{equation}
    \frac{H}{H_{max}} = \frac{1}{1+\frac{3}{2}|1+w_{DE}|H_{max}(t_{max}-t)} \label{H_t} \,.
\end{equation}

\begin{figure}[htpb]
    \centering
    \includegraphics[width=0.9\textwidth]{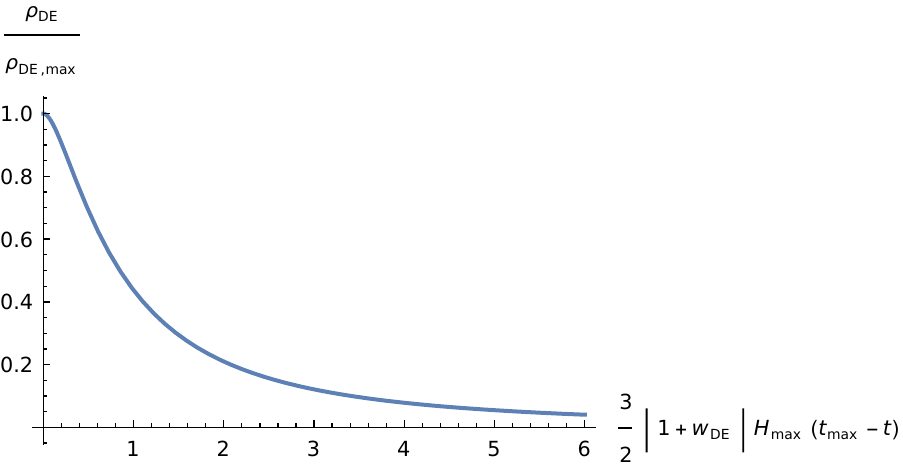}
    \caption{The plot of $\displaystyle\frac{\rho_{DE}}{\rho_{DE,max}}$ as a function of $t_{max}-t$. $\rho_{DE} \rightarrow \rho_{DE,max}$ as $t \rightarrow t_{max}$.}
    \label{fig:R_t}
\end{figure}

\begin{figure}[htpb]
    \centering
    \includegraphics[width=0.9\textwidth]{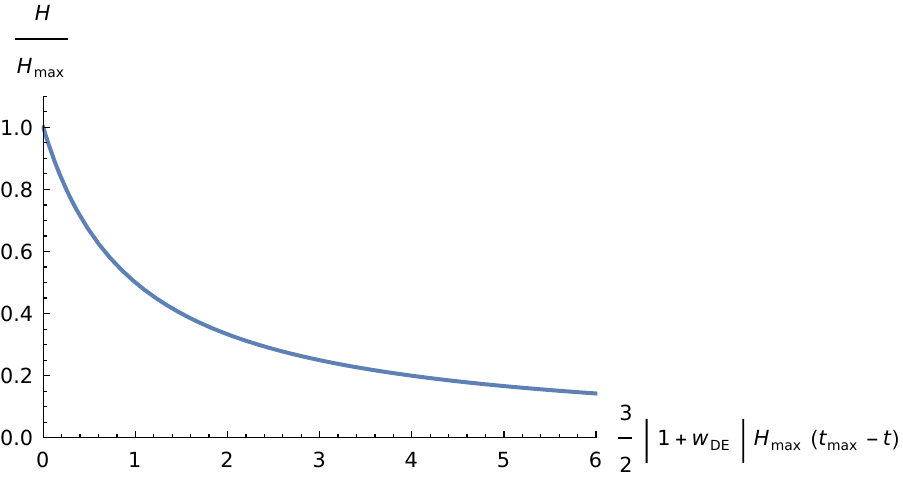}
    \caption{The plot of $\displaystyle\frac{H}{H_{max}}$ as a function of $t_{max}-t$. $H \rightarrow H_{max}$ as $t \rightarrow t_{max}$.}
    \label{fig:H_t}
\end{figure}

From Figure \eqref{fig:R_t} and Figure \eqref{fig:H_t}, one can see that the energy density and Hubble rate do not reach infinite values in a finite amount of time, as it would happen in the case of a Big Rip scenario. On the contrary, they reach the maximum values $\rho_{max}$ and $H_{max}$ respectively as $t \rightarrow t_{max}$ and therefore the Big Rip singularity is avoided. This happens thanks to the ever increasing rate of Schwinger pair-production that an observer inside a causal horizon of radius $r_H \sim H^{-1}$ would observe as the Hubble rate increases.

\subsection{Horizon entropy in the far-future of the universe} \label{subsec:horizon_entropy}

It is also important to have a qualitative understanding of the evolution of the horizon entropy in the scenario discussed in this paper. Its calculation is quite straightforward. Using the analogy between the thermodynamics of black holes and cosmological horizons, the Generalised Second Law (GSL) for black holes \cite{GSL_BH}, \cite{GSL_BH_new_1}, \cite{GSL_BH_new_2} extended to de Sitter horizons is \cite{gibbons_hawking_1}, \cite{gibbons_hawking_2}

\begin{equation}
    \Delta(S_{outside} + S_{H}) \geq  0 \label{deltaS} \,,
\end{equation}

\begin{equation}
    S_{H} \propto A_H \propto r_{H}^2 \propto H^{-2} \label{S_H} \,,
\end{equation}

where $S_H$ is the horizon entropy, $S_{outside}$ is the entropy outside the horizon, $A_H$ is the area of the de Sitter horizon and $r_{H}$ is the Hubble radius\footnote{``Outside" for the case of the black holes is ``inside" for the case of de Sitter space, and in general for the case of cosmological horizons.}. 

The relations \eqref{deltaS} and \eqref{S_H} have been proven to hold true also in the case of accelerated horizons and are independent of whether the horizon area $A_H$ increases or decreases, as long as the rate of increase of the entropy outside the horizon $S_{outside}$ outweighs the rate of decrease of the horizon entropy $S_{H}$ \cite{horizon_entropy}.

In the case analysed in this paper, where the universe is dominated by the phantom fluid, the horizon entropy decreases because of the increase of the Hubble rate, until it reaches asympotically a minimum constant value $S_{H,min}$ as $t \rightarrow t_{max}$ which is 

\begin{equation}
    S_{H,min} \propto H_{max}^{-2} \propto C_S \label{S_H_min} \,.
\end{equation}

One observes that the entropy does not reach a zero value thanks to the particle-pair production because of the Schwinger effect. When $S_H$ reaches its minimum value $S_{H,min} \sim \mathcal{O}{(C_S)}$ in the far-future, the universe enters into a de Sitter phase of constant and high Hubble rate $H_{max}$ and energy density $\rho_{max}$. The fact that the horizon entropy decreases is not a problem as long as the causal patch of an observer is not an isolated system and the rate of increase of the entropy outside the horizon outweighs the rate of decrease of the horizon entropy, as discussed above\footnote{An idea has been recently introduced: the central dogma about cosmological horizons \cite{susskind}. This idea is an extension of the central dogma about black holes \cite{maldacena} to cosmological horizons and considers that every causal patch is supposed to be an isolated system. If this conjecture would hold true, the presence of a phantom substance would violate the second law of thermodynamics, which is one of the most sacred laws of Physics, therefore the whole discussion in this paper would have to be abandoned. However, this dogma is based on holographic arguments, which silently imply the NEC, which by definition is violated in the case of a phantom fluid. This discussion is also done in \cite{ijjas_steinhardt}, but for bouncing models. Therefore, the whole discussion in this paper would be ruled out just by bias and not by any independent arguments.}.

\begin{figure}[htpb]
    \centering
    \includegraphics[width=0.9\textwidth]{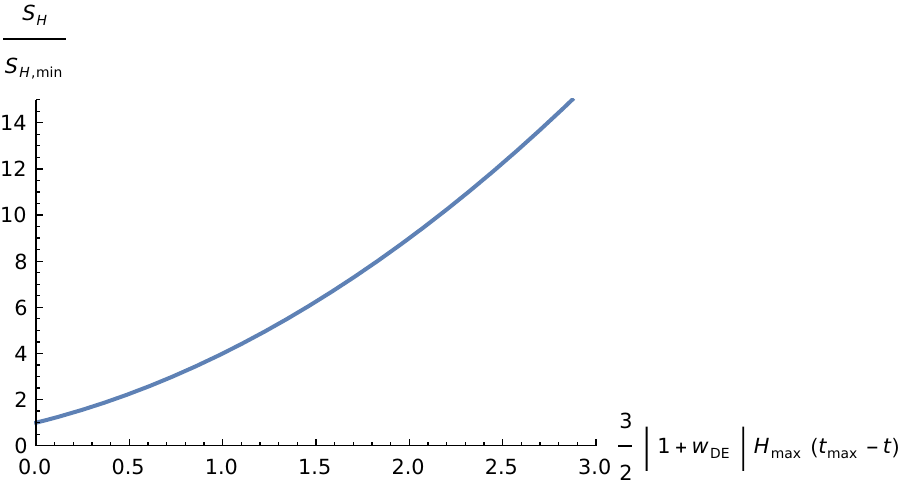}
    \caption{The plot of $\displaystyle\frac{S_H}{S_{H,min}}$ as a function of $t_{max}-t$. $S_H \rightarrow S_{H, min}$ as $t \rightarrow t_{max}$.}
    \label{fig:SH_t}
\end{figure}

Using \eqref{H_2_final} and \eqref{S_H_min} results in the following evolution of the horizon entropy \eqref{S_H} (Figure \eqref{fig:SH_t})

\begin{equation}
    \frac{S_{H}}{S_{H,min}} = \Big(1+\frac{3}{2}|1+w_{DE}|H_{max}(t_{max}-t)\Big)^2 \label{S_H_final} \,.
\end{equation}

\subsection{Cyclic/Periodic model of the universe?}

\begin{figure}[H]
    \centering
    \includegraphics[width=0.9\textwidth]{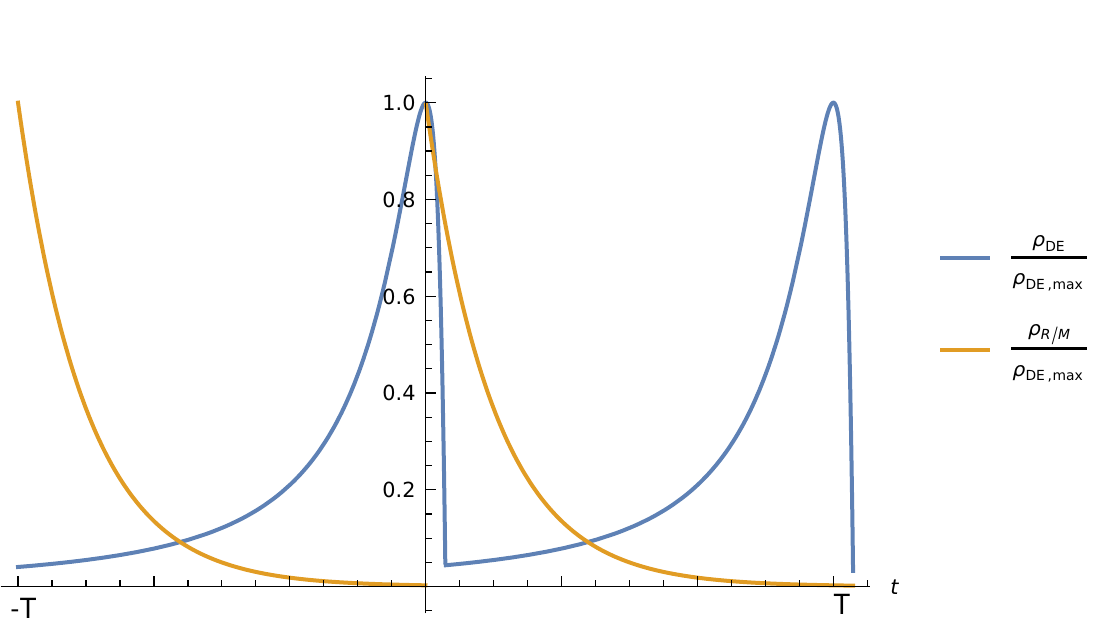}
    \caption{The plots of the phantom energy density $\rho_{DE}$ and the radiation/matter energy density $\rho_{R/M}$ as a function of time $t$ in two cycles, each of period T, according to the model described in this work. One can see how in each period the phantom energy density eventually dominates over radiation/matter. It never reaches infinity, but rather a constant and high value $\rho_{DE,max}$, thanks to the backreaction from the gravitational Schwinger pair-production. Then the universe passes smoothly to an inflationary phase with $\rho_{DE} \sim \rho_{DE,max}$ until its decay refills the universe with radiation and matter. Therefore a continuous cycle can be assumed where superacceleration, followed by inflation, followed by reheating, followed by matter/radiation domination, followed by phantom substance domination takes place periodically in the history of the universe. In this graph, the same scaling for matter and radiation is assumed for simplicity. In general, each cycle can have a different value of period T, which can be defined by the specific characteristics of each cycle of the universe.}
    \label{fig:rho_t_final}
\end{figure}

The combination of the results of paragraphs \eqref{subsec:schwinger_pairs} and \eqref{subsec:horizon_entropy} leads to the following conclusion: If the universe is dominated by a phantom fluid with $w_{DE}<-1$ and if the gravitational Schwinger pair-production takes place, the Big Rip singularity could be avoided in the far-future. In particular, because the $\rho = \rho_{DE,max}$ and $H = H_{max}$ are solutions to the equations \eqref{H_2} and \eqref{continuity_eq_DE_source}, the energy density and the Hubble rate would remain constant after they reach these values in the far-future, according to the analysis in this paper, and the universe could start a phase of de Sitter inflation. Then, by taking one more step of speculation, the phantom substance could decay through an unknown hypothetical mechanism, which would lead to the reheating of the universe and to a new Big Bang; thus the beginning of a new cycle of the universe. Of course, the specific characteristics of the decay of the phantom fluid to matter and radiation would have to be understood and this is beyond the scope of this paper \footnote{An idea similar to the time-periodic (cyclic) universe, discussed in this work, has already been suggested in \cite{senatore2006}, but there in the context of effective field theories. However, in this paper the author uses additional entropic arguments and a mechanism of stopping smoothly the ever increasing energy density and Hubble rate through the gravitational Schwinger effect, before the decay of the phantom fluid to radiation/matter in the next cycle.}.This idea can be seen graphically in Figure \eqref{fig:rho_t_final}. 

\section{Conclusions}

In this work, the behaviour of a universe dominated by phantom energy with generic equation of state parameter $w<-1$ is analysed. Working solely in the context of general relativity and assuming no instabilities caused by the violation of NEC, it is found that the Big Rip singularity could potentially be avoided in the far-future because of the gravitational Schwinger pair-production. The universe would reach a high but constant Hubble rate $H_{max}$ and energy density $\rho_{max}$, passing to a de Sitter inflationary phase in a finite amount of time $t_{max}$. As a final step it is assumed that the phantom substance could decay to matter/radiation and reheat the universe until the first would dominate again, leading thus to a cyclic/periodic model of the universe. This situation could in theory be repeated an infinite amount of times, unless some other process stops it or changes it. This cyclic/periodic model of the universe is highly speculative and the specific dynamics of the phantom substance decay to radiation and matter are left to be investigated in future works. Even more importantly, it should be clarified from future observations whether dark energy is a phantom fluid, a cosmological constant or something completely different and unexpected.

\section*{Acknowledgements}
The author would like to thank Hitoshi Murayama and Yasunori Nomura for useful discussions, and especially Simone Ferraro for the useful guidance, valuable inputs and his support throughout this work.

\end{document}